\documentclass[onecolumn]{openjournal}

\usepackage[T1]{fontenc}
\usepackage[utf8]{inputenc}
\usepackage{lmodern}
\usepackage{microtype}

\usepackage{amsmath,amssymb}
\usepackage{graphicx}
\usepackage{booktabs}
\usepackage{adjustbox}
\usepackage{enumitem}
\usepackage{xcolor}
\usepackage{hyperref}
\usepackage{url}

\hypersetup{
    colorlinks=true,
    linkcolor=blue!50!black,
    citecolor=blue!50!black,
    urlcolor=blue!50!black,
    pdftitle={Native-Opacity Sensitivity of a Fixed Delta Cephei MESA-RSP Pulsation Model},
    pdfauthor={Zuhoor Elahi, Wafa Gull, Christopher Sirola}
}

\graphicspath{{./}{opacity/}}
\newcommand{\dcep}{\ensuremath{\delta} Cephei}

\newcommand{\Prsp}{\ensuremath{P_{\rm RSP}}}

\newcommand{\DMag}{\ensuremath{\Delta{\rm Mag}}}
\newcommand{\DR}{\ensuremath{\Delta R}}
\newcommand{\GREKM}{\texorpdfstring{\ensuremath{\text{\texttt{rsp\_GREKM}}}}{rsp_GREKM}}

\newcommand{\DVphot}{\ensuremath{\Delta v_{\rm phot,proxy}}}

\begin{document}

\title{Native-Opacity Sensitivity of a Fixed Delta Cephei MESA-RSP Pulsation Model}

\author{Zuhoor Elahi}
\email{zuhoor.elahi@usm.edu}
\affiliation{Department of Physics and Astronomy, University of Southern Mississippi, Hattiesburg, MS, USA}
\affiliation{Department of Physics, University of Karachi, Karachi, Pakistan}

\author{Wafa Gull}
\affiliation{Department of Physics and Astronomy, University of Southern Mississippi, Hattiesburg, MS, USA}

\author{Christopher Sirola}
\affiliation{Department of Physics and Astronomy, University of Southern Mississippi, Hattiesburg, MS, USA}

\begin{abstract}
Radiative opacity is one of the central microphysical inputs controlling the thermal response of Cepheid envelopes and the driving or damping of radial pulsations.  We present a controlled opacity-sensitivity experiment for a fixed \dcep{} nonlinear radial pulsation model computed with the MESA Radial Stellar Pulsation module.  The stellar and pulsation parameters are held fixed at \(M=5.0\,M_\odot\), \(T_{\rm eff}=6050\,{\rm K}\), \(L=2360\,L_\odot\), \(X=0.73\), \(Z=0.007\), and \texttt{RSP\_alfam}=0.425, while the high-temperature opacity source is varied among native MESA opacity configurations: OPAL-A09, OP-A09, and OPLIB-AGSS09.  This parameter set is inherited from an earlier OPAL-A09-based \dcep{} calibration sequence, so the present study is a differential opacity-sensitivity test about that reference model rather than an independent recalibration for each opacity family.  The low-temperature opacity prefix is held fixed; the C/O Type-2 prefix is also unchanged but is inactive for these hydrogen-rich envelopes under the MESA Type-2 blending criteria.  Verification integrations were performed at 20, 100, and 300 pulsation cycles, followed by photo-restarted continuations to 500 cycles.  Within this fixed OPAL-calibrated configuration, OPAL-A09 gives the closest period agreement, \(P_{\rm RSP}=5.366986\,{\rm d}\), only about 39 s longer than \(P_{\rm obs}=5.366531\,{\rm d}\).  OP-A09 gives the largest amplitude-growth diagnostics, with \(\Delta{\rm Mag}=0.037307\) and \(\Delta R=0.293677\), corresponding to increases of 42.5\% and 43.9\% relative to OPAL-A09.  OPLIB-AGSS09 gives a systematically longer period, \(P_{\rm RSP}=5.403926\,{\rm d}\), with more modest amplitude-growth changes.  The same ordering is reflected in the MESA history-column diagnostic \GREKM{}, defined by the MESA defaults as the fractional growth of kinetic energy per pulsation period.  These results show that native opacity choice measurably affects period matching, pulsation growth diagnostics, and nonlinear amplitude growth in this fixed \dcep{} model.  Because the reference parameters were calibrated in an OPAL-based sequence, the period comparison does not establish that OPAL would remain preferred after opacity-specific retuning.  However, the tested opacity choices do not by themselves resolve the observed-amplitude discrepancy.
\end{abstract}

\keywords{Cepheid variables, Delta Cephei, stellar pulsation, radiative opacity, MESA, MESA-RSP, OPAL, OP, OPLIB}

\section{Introduction}
\label{sec:introduction}

Classical Cepheids are radially pulsating evolved stars whose periods, luminosities, and light-curve shapes are central to stellar astrophysics and distance-scale work.  The prototype \dcep{} is a useful benchmark because it is nearby, well observed, and has a precisely known pulsation period.  In numerical Cepheid modeling, reproducing the observed period is a necessary but not sufficient test.  A model can match the period while still failing to reproduce the observed optical amplitude or the detailed morphology of the light curve.  This motivates controlled tests of the physical and numerical inputs that affect the nonlinear pulsation response.

Radiative opacity is a particularly important input because Cepheid pulsation is driven by the heat-engine behavior of partial-ionization zones.  In the \(\kappa\)-mechanism, compression of an ionization zone can increase the ability of the gas to trap radiation, delaying radiative energy leakage and feeding mechanical work into the pulsation at the proper phase.  The resulting driving or damping depends on the opacity profile, the local thermal timescale, and the envelope structure.  Therefore, even when global quantities such as mass, luminosity, and effective temperature are held fixed, changes in the opacity tables can alter the pulsation period, growth rate, and amplitude development.

The importance of opacity for classical pulsators has long been recognized.  The modern stellar-opacity problem was emphasized in the context of Cepheid pulsation by Simon \citep{simon1982}, and subsequent OPAL opacity calculations changed the interpretation of several stellar-pulsation problems \citep{iglesiasrogers1991,iglesiasrogers1996}.  The Opacity Project and later Los Alamos OPLIB calculations provide alternative high-temperature opacity inputs that are now available in stellar-evolution and pulsation calculations \citep{badnell2005,seaton2005,colgan2016,farag2024}.  Comparative Cepheid calculations with OP and OPAL opacities have also been discussed in earlier pulsation studies \citep{moskalik1992,kanbur1994}.  These works motivate a focused question in the present MESA-RSP context: when all other model parameters are held fixed, how much does the native MESA opacity choice change the nonlinear pulsation response of a \dcep{} model?  Recent work by \citet{farag2026cepheid} has also demonstrated self-consistent nonlinear classical-Cepheid pulsations during MESA stellar evolution and compared matched MESA-star and MESA-RSP envelope calculations, providing additional context for the RSP envelope approach used here.

This paper addresses that question through a controlled numerical experiment.  We hold the stellar and RSP parameters fixed and vary only the high-temperature opacity configuration among three native MESA choices: OPAL-A09, OP-A09, and OPLIB-AGSS09.  The fixed parameter set was inherited from the preceding OPAL-A09-based period and amplitude-control sequence \citep{elahi2026nonlinear}.  It is therefore an OPAL-calibrated reference configuration, not an opacity-neutral global best fit.  The low-temperature opacity prefix and the C/O-dependent opacity prefix are kept fixed.  This setup does not attempt to recalibrate the best possible \dcep{} model for each opacity table.  Instead, it isolates the differential sensitivity of one calibrated RSP setup to opacity choice.  Accordingly, a close OPAL period match in this experiment should not be interpreted as a general ranking of opacity tables; modest retuning of stellar or convection parameters could plausibly recover a comparable period match for another opacity choice.

The main questions are:
\begin{enumerate}[label=(\roman*)]
    \item Does the high-temperature opacity choice change the final RSP period in a fixed \dcep{} model?
    \item Does opacity choice change the MESA-RSP amplitude-growth diagnostics \DMag{} and \DR?
    \item Does the MESA history-column diagnostic \GREKM{} support the same ordering as the amplitude-growth diagnostics?
    \item Can changing among native opacity tables alone remove the discrepancy between the model amplitude and the observed optical amplitude of \dcep?
\end{enumerate}

\section{Physical Basis of the Opacity Sensitivity}
\label{sec:physics}

\subsection{Opacity, thermal structure, and pulsation driving}

In a radiative stellar layer, the radiative flux depends inversely on the Rosseland mean opacity.  In the diffusion approximation,
\begin{equation}
    F_{\rm rad} = -\frac{4acT^3}{3\kappa\rho}\frac{dT}{dr},
    \label{eq:radiative_flux}
\end{equation}
where \(a\) is the radiation constant, \(c\) is the speed of light, \(T\) is temperature, \(\rho\) is density, and \(\kappa\) is opacity.  A different opacity table therefore changes the relation between the local temperature gradient and the outward radiative flux.  In a pulsating envelope, this changes how quickly thermal energy escapes during compression and expansion.

The radiative temperature gradient is often written schematically as
\begin{equation}
    \nabla_{\rm rad} \propto \frac{\kappa L P}{M T^4},
    \label{eq:radgrad}
\end{equation}
where \(L\), \(P\), \(M\), and \(T\) denote local luminosity, pressure, enclosed mass, and temperature.  This standard radiative-gradient scaling follows from the stellar-structure equations for radiative energy transport \citep{kippenhahn2012}.  Thus, a change in \(\kappa(T,\rho,X,Z)\) can affect the envelope stratification even when the global stellar parameters are unchanged.

For classical Cepheids, the most important driving regions are partial-ionization zones, especially helium ionization zones.  During compression, an ionization zone can temporarily store thermal energy rather than immediately radiating it away.  If the opacity response has the correct phase relation with the pulsation, the trapped heat increases pressure during expansion and adds energy to the pulsation.  If the phase relation is unfavorable, the same region can contribute to damping.  This is why the opacity profile is not merely a background input; it is part of the physical mechanism that determines pulsation growth.

\subsection{Period and amplitude diagnostics}

The approximate scale of a radial pulsation period is set by the dynamical timescale.  A standard period--mean-density relation is
\begin{equation}
    P \simeq Q\left(\frac{R^3}{GM}\right)^{1/2},
    \label{eq:period_mean_density}
\end{equation}
where \(Q\) is the pulsation constant, \(R\) is the stellar radius, \(M\) is the stellar mass, and \(G\) is the gravitational constant \citep{cox1980}.  The nonlinear MESA-RSP period is not imposed directly by Eq.~\eqref{eq:period_mean_density}; nevertheless, the relation provides the physical intuition that changes in envelope structure, effective radius, and pulsation constant can shift the computed period.

The present work uses four main MESA-RSP diagnostics:
\begin{align}
    \Prsp &\equiv \text{the RSP pulsation period}, \\
    \DMag &\equiv \text{the internal MESA-RSP magnitude-amplitude diagnostic}, \\
    \DR &\equiv \text{the internal MESA-RSP radius-amplitude diagnostic}, \\
    \GREKM{} &\equiv \text{the MESA-RSP fractional growth of kinetic energy per pulsation period}.
\end{align}
For an additional velocity comparison, we additionally compute \DVphot{} as the peak-to-peak finite-difference derivative of the MESA history quantity \texttt{photosphere\_r} over the final complete pulsation cycle.  This is a photospheric radial-velocity proxy, not a synthetic spectroscopic line-of-sight velocity.
The quantities \DMag{} and \DR{} are used to compare nonlinear amplitude growth among opacity cases.  The quantity \DMag{} is not a calibrated Johnson-\(V\) amplitude; a calibrated observed-band light curve requires a separate synthetic-photometry or bolometric-correction step.  The quantity \GREKM{} is a MESA history-column output defined in the MESA r24.08.1 default history-column list as the fractional growth of kinetic energy per pulsation period, described there as a ``nonlinear growth rate'' and linked to Eq.~5 of the MESA V paper \citep{mesa_history_columns_2024,paxton2019}.  It is included only as a secondary internal diagnostic for relative comparison among the three opacity cases and is not an observed quantity.

An independent peer-reviewed observational compilation by \citet{dasilva2022} gives for \dcep{} a period of \(5.3663\,{\rm d}\), a Johnson-\(V\) peak-to-peak amplitude of \(0.84\pm0.04\,{\rm mag}\), and a peak-to-peak spectroscopic radial-velocity amplitude of \(38.9\pm4.9\,{\rm km\,s^{-1}}\).  The more precise period and optical-amplitude values used below are adopted from companion observational analyses of \dcep{}.  The AAVSO Johnson-\(V\) Fourier-template study provides the observed light-curve morphology and amplitude scale, while the semi-empirical reconstruction connects the photometry with radial velocities, temperature constraints, and SPIPS-based phase curves \citep{elahi2026fourier,elahi2026spips}:
\begin{equation}
    P_{\rm obs}=5.366531\,{\rm d}
    \label{eq:pobs}
\end{equation}
and
\begin{equation}
    \Delta V_{\rm obs}\simeq 0.8390\,{\rm mag}.
    \label{eq:dvobs}
\end{equation}
The independent values from \citet{dasilva2022} are consistent with these adopted benchmarks, and the small numerical differences have no effect on the opacity-ordering conclusions.  The comparison to Eq.~\eqref{eq:dvobs} is qualitative in this paper because the RSP \DMag{} diagnostic is not the final calibrated Johnson-\(V\) amplitude.

\section{Model Setup and Numerical Procedure}
\label{sec:methods}

\subsection{Fixed MESA-RSP model}

All calculations use the MESA-RSP framework implemented in MESA \citep{paxton2011,paxton2013,paxton2015,paxton2018,paxton2019}.  The model is a fixed \dcep{} RSP setup with a single mass, luminosity, effective temperature, composition, and turbulent-viscosity parameter.  The purpose is not to search a broad grid for the best possible model.  The purpose is to ask how the same model responds when only the high-temperature opacity table is changed.

Table~\ref{tab:model_setup} gives the fixed stellar and pulsation parameters.  These values are held constant for all opacity cases.

\begin{table}[t]
\centering
\caption{Fixed MESA-RSP parameters used in the opacity-sensitivity calculations.}
\label{tab:model_setup}
\begin{tabular}{ll}
\toprule
Quantity & Value \\
\midrule
Mass & \(5.0\,M_\odot\) \\
Effective temperature & \(6050\,{\rm K}\) \\
Luminosity & \(2360\,L_\odot\) \\
Hydrogen mass fraction & \(X=0.73\) \\
Metallicity & \(Z=0.007\) \\
RSP turbulent-viscosity parameter & \texttt{RSP\_alfam = 0.425} \\
RSP radiative-loss parameter & \texttt{RSP\_gammar = 0.0} \\
Reference opacity used in preceding calibration & OPAL-A09 \\
Observed period used for comparison & \(P_{\rm obs}=5.366531\,{\rm d}\) \\
Observed optical amplitude used for context & \(\Delta V_{\rm obs}\simeq0.8390\,{\rm mag}\) \\
Observed spectroscopic radial-velocity amplitude & \(\Delta RV_{\rm obs}=38.9\pm4.9\,{\rm km\,s^{-1}}\) \citep{dasilva2022} \\
\bottomrule
\end{tabular}
\end{table}

The fixed values in Table~\ref{tab:model_setup} were inherited from the preceding OPAL-A09 MESA-RSP calibration sequence \citep{elahi2026nonlinear}.  That sequence first tuned the global stellar model to the observed period and subsequently explored the eddy-viscous parameter \texttt{RSP\_alfam}; the \(L=2360\,L_\odot\), \texttt{RSP\_alfam}=0.425, 500-cycle calculation was retained as the clean amplitude-enhanced reference.  The present opacity experiment therefore begins from an explicitly OPAL-calibrated parameter set.  Because the goal here is a one-variable-at-a-time differential test, no retuning of mass, luminosity, temperature, composition, or RSP turbulent parameters is performed for OP or OPLIB.  Such opacity-specific recalibration could alter which table gives the closest period match and is left to a broader grid study.

\subsection{Origin of the fixed input parameters and observational context}
\label{sec:parameter_provenance}

The adopted values were not obtained by simultaneously fitting all of the independent observables of \dcep{}. The model was developed in the earlier OPAL-A09-based period-calibration and nonlinear amplitude-control sequence \citep{elahi2026nonlinear}. A mass near five solar masses was adopted as a plausible starting point for the classical Cepheid, while the input effective temperature and luminosity were adjusted within that modeling workflow to obtain a useful pulsation-period reference; the viscosity coefficient was then selected from an amplitude-control sequence. The resulting choices were preserved exactly in the present experiment, rather than independently reoptimized for OP and OPLIB. We therefore distinguish the fixed \emph{model inputs} from the independently constrained stellar properties listed in Table~\ref{tab:observational_context}.

\begin{table}[htbp]
\centering
\caption{Context for the adopted fixed inputs. Literature values characterize \dcep{} and are not error bars on the RSP inputs. Mass estimates are model dependent; the temperature interval reflects variation with pulsation phase. The luminosity is approximate and does not come with a quoted uncertainty in the cited compilation. The measured [Fe/H] does not uniquely determine the mass fraction \(Z\) without an assumed abundance mixture and helium/hydrogen fraction.}
\label{tab:observational_context}
\small
\begin{tabular}{@{}p{0.17\linewidth}p{0.17\linewidth}p{0.58\linewidth}@{}}
\toprule
Quantity & Fixed RSP input & Independent literature constraint / status \\
\midrule
Mass & \(5.0\,M_\odot\) & Pulsation: \(4.5\pm0.3\,M_\odot\); evolutionary: \(5.7\pm0.5\,M_\odot\) \citep{matthews2012,engle2014} \\
Effective temperature & \(6050\,{\rm K}\) & Phase range \(\simeq5500\)--\(6600\,{\rm K}\) \citep{engle2014}; mean \(5905\pm22\,{\rm K}\) \citep{dasilva2022} \\
Luminosity & \(2360\,L_\odot\) & Mean luminosity \(\sim2000\,L_\odot\) (approximate) \citep{matthews2012,engle2014} \\
Hydrogen fraction & \(X=0.73\) & No direct \dcep{} determination adopted; fixed composition assumption \\
Metal fraction & \(Z=0.007\) & Spectroscopic \([\mathrm{Fe/H}]=+0.05\pm0.02\,{\rm dex}\) \citep{dasilva2022} \\
\bottomrule
\end{tabular}
\end{table}

The two published mass values in Table~\ref{tab:observational_context} are distinct pulsation- and evolutionary-model estimates, not direct dynamical measurements or a single combined confidence interval. The value \(M=5.0\,M_\odot\) lies between them. In contrast, the adopted \(T_{\rm eff}=6050\,{\rm K}\) is within the observed phase-dependent temperature excursion but is higher than the reported mean of \(5905\pm22\,{\rm K}\); the \(\pm22\,{\rm K}\) is the reported uncertainty on the spectroscopic mean and must not be interpreted as the full phase range. Similarly, \(L=2360\,L_\odot\) exceeds the approximate published mean \(\sim2000\,L_\odot\), so we do not claim that the chosen luminosity is an independent best-fit determination. The adopted \(X=0.73\) is a near-solar hydrogen mass-fraction assumption (for comparison, Asplund et al. \citep{asplund2009} quote solar photospheric \(X_\odot=0.7381\)), not a measured abundance for this star; the corresponding uniform helium fraction is \(Y=1-X-Z=0.263\).

The adopted metallicity deserves an explicit caution. Da Silva et al. \citep{dasilva2022} obtain a near-solar spectroscopic abundance \([\mathrm{Fe/H}]=+0.05\pm0.02\,{\rm dex}\), with an approximately \(0.05\,{\rm dex}\) scatter across their individual abundance measurements. For orientation only, the solar-scaled relation \([\mathrm{M/H}]\simeq\log_{10}[(Z/X)/(Z/X)_\odot]\), using \((Z/X)_\odot=0.0181\) from \citet{asplund2009}, gives about \(-0.28\,{\rm dex}\) for the model \(X=0.73\), \(Z=0.007\). This approximate comparison assumes a solar-scaled mixture and \([\mathrm{M/H}]\simeq[\mathrm{Fe/H}]\); it is not a direct measurement of the star's \(Z\). It shows that \(Z=0.007\) should \emph{not} be described as spectroscopically calibrated to \dcep{}. Its use here is inherited from the period-tuned RSP reference and is held fixed to isolate the opacity-table effect. A near-solar-abundance, opacity-specific stellar-parameter grid would be needed for a more physically constrained fit to the star.

\subsection{Envelope construction and initialization}

The three calculations use the MESA r24.08.1 RSP model builder with a single global envelope composition, \(X=0.73\) and \(Z=0.007\), rather than an evolved abundance profile.  The RSP mesh and inner-boundary controls were not overridden, so the r24.08.1 defaults apply: \texttt{RSP\_nz}=150, \texttt{RSP\_nz\_outer}=40, \texttt{RSP\_T\_anchor}=11,000 K, and \texttt{RSP\_T\_inner}=\(2.0\times10^6\) K \citep{mesa_controls_2024}.  The run logs independently report 150 zones in every case.  The same defaults initialize the pulsation with a scaled linear eigenvector having a surface-velocity kick of \(0.1\,{\rm km\,s^{-1}}\); the first- and second-overtone fractions are zero, so the kick is purely fundamental-mode \citep{mesa_controls_2024}.

Each opacity case was constructed directly with \texttt{create\_RSP\_model = .true.} and its own high-temperature \texttt{kap\_file\_prefix}.  OP and OPLIB were therefore not initialized from an OPAL model and were not dynamically ``switched'' away from OPAL.  The different opacity tables produce small differences in the relaxed static-envelope solution even though the same mesh size and temperature-boundary controls are used.  Table~\ref{tab:envelope_structure} lists the initial-envelope diagnostics reported by the run logs.  The inner-boundary mass and radius remain fixed during each subsequent nonlinear integration and its 300-to-500-cycle restart.

\begin{table}[t]
\centering
\caption{Initial MESA-RSP envelope diagnostics for the three opacity cases.  The common inner-temperature target is \texttt{RSP\_T\_inner}=\(2.0\times10^6\) K; \(T_{\rm in,cell}\) is the innermost zone-center temperature reported at construction.}
\label{tab:envelope_structure}
\begin{adjustbox}{max width=\textwidth}
\begin{tabular}{lrrrrr}
\toprule
Opacity case & Zones & \(T_{\rm in,cell}\) (K) & \(R_{\rm in}/R_\odot\) & \(M_{\rm in}/M_\odot\) & Envelope mass (\(M_\odot\)) \\
\midrule
OPAL-A09 & 150 & 2.001208\(\times10^6\) & 5.6940 & 3.85595 & 1.14405 \\
OP-A09 & 150 & 2.001216\(\times10^6\) & 5.6879 & 3.84368 & 1.15632 \\
OPLIB-AGSS09 & 150 & 2.000248\(\times10^6\) & 5.6815 & 3.89970 & 1.10030 \\
\bottomrule
\end{tabular}
\end{adjustbox}
\end{table}

The \texttt{kap\_CO\_prefix='a09\_co'} setting is retained identically in all inlists for reproducibility, but it does not affect this comparison.  In MESA r24.08.1, C/O-enhanced Type-2 opacities are fully disabled in a cell when \(X\ge10^{-3}\), and they are also suppressed when \(Z-Z_{\rm base}\) is small \citep{mesa_kap_defaults_2024}.  Here \(X=0.73\) and \(Z=Z_{\rm base}=0.007\), so the Type-2 contribution is zero and the C/O prefix is inactive.  The physically relevant opacity change is therefore the hydrogen-rich high-temperature Type-1 table selected by \texttt{kap\_file\_prefix}.

\subsection{Opacity configurations}

The three opacity configurations are listed in Table~\ref{tab:opacity_configurations}.  The high-temperature opacity prefix is varied among OPAL-A09, OP-A09, and OPLIB-AGSS09.  The low-temperature opacity prefix and C/O-dependent opacity prefix are fixed as \texttt{lowT\_fa05\_a09p} and \texttt{a09\_co}, respectively.  The latter is listed to document the complete inlist configuration; as noted above, the C/O Type-2 table has zero weight for the present hydrogen-rich, \(Z=Z_{\rm base}\) envelopes.

\begin{table}[t]
\centering
\caption{Native MESA opacity configurations tested in the controlled comparison.}
\label{tab:opacity_configurations}
\begin{adjustbox}{max width=\textwidth}
\begin{tabular}{lllll}
\toprule
Case & High-temperature prefix & Low-temperature prefix & C/O prefix & Role \\
\midrule
OPAL-A09 & \texttt{a09} & \texttt{lowT\_fa05\_a09p} & \texttt{a09\_co} & Reference case \\
OP-A09 & \texttt{OP\_a09\_nans\_removed\_by\_hand} & \texttt{lowT\_fa05\_a09p} & \texttt{a09\_co} & OP comparison \\
OPLIB-AGSS09 & \texttt{oplib\_agss09} & \texttt{lowT\_fa05\_a09p} & \texttt{a09\_co} & OPLIB comparison \\
\bottomrule
\end{tabular}
\end{adjustbox}
\end{table}

The corresponding MESA opacity control block has the form
\begin{verbatim}
&kap
   Zbase = 0.007d0
   kap_file_prefix = '<case-dependent high-temperature prefix>'
   kap_lowT_prefix = 'lowT_fa05_a09p'
   kap_CO_prefix   = 'a09_co'
/
\end{verbatim}
The MESA opacity controls were used as native MESA table selections rather than as externally modified opacity tables \citep{mesa2024docs}.

\subsection{Run sequence}

The calculations were performed in a staged sequence.  The 20-cycle runs verified that each opacity table selection produced a valid RSP calculation.  The 100- and 300-cycle runs tested the persistence of the opacity-dependent trends.  The final production comparison uses 500-cycle histories, obtained by continuing from the final 300-cycle state through a MESA photo restart.  This procedure preserves the evolved nonlinear pulsation state and avoids repeating the initial transient.  Each 300-to-500-cycle continuation was restarted from that same opacity case's own 300-cycle photo, with the same case-specific \texttt{kap\_file\_prefix}; no opacity table was changed during a restart.

The requested stopping condition was simply completion of 500 pulsation periods, set by \texttt{RSP\_max\_num\_periods = 500}.  The logs terminate with the MESA message that the maximum number of periods was reached.  It was not a convergence criterion on period, amplitude, or \GREKM{}.  The 20-, 100-, 300-, and 500-cycle summaries are retained in the paper because they show that the opacity-dependent trends are not a single-cycle artifact.  However, the physical interpretation is based primarily on the 500-cycle comparison.

\subsection{Scope of included and excluded calculations}

The final physical interpretation is restricted to native MESA opacity choices.  OPAS and other custom-opacity options are not used for final numerical conclusions in this paper.  A custom opacity implementation must first pass checks of table format, composition consistency, C/O coverage, thermodynamic-domain coverage, interpolation behavior, and recovery of a reference case.  Without those checks, a custom-opacity calculation could reflect a table-format or interpolation artifact rather than a physical opacity effect.  Custom-opacity work is therefore reserved for a separate validation study.

\section{Results}
\label{sec:results}

\subsection{Final 500-cycle period response}

\begin{figure}[t]
    \centering
    \includegraphics[width=0.86\linewidth]{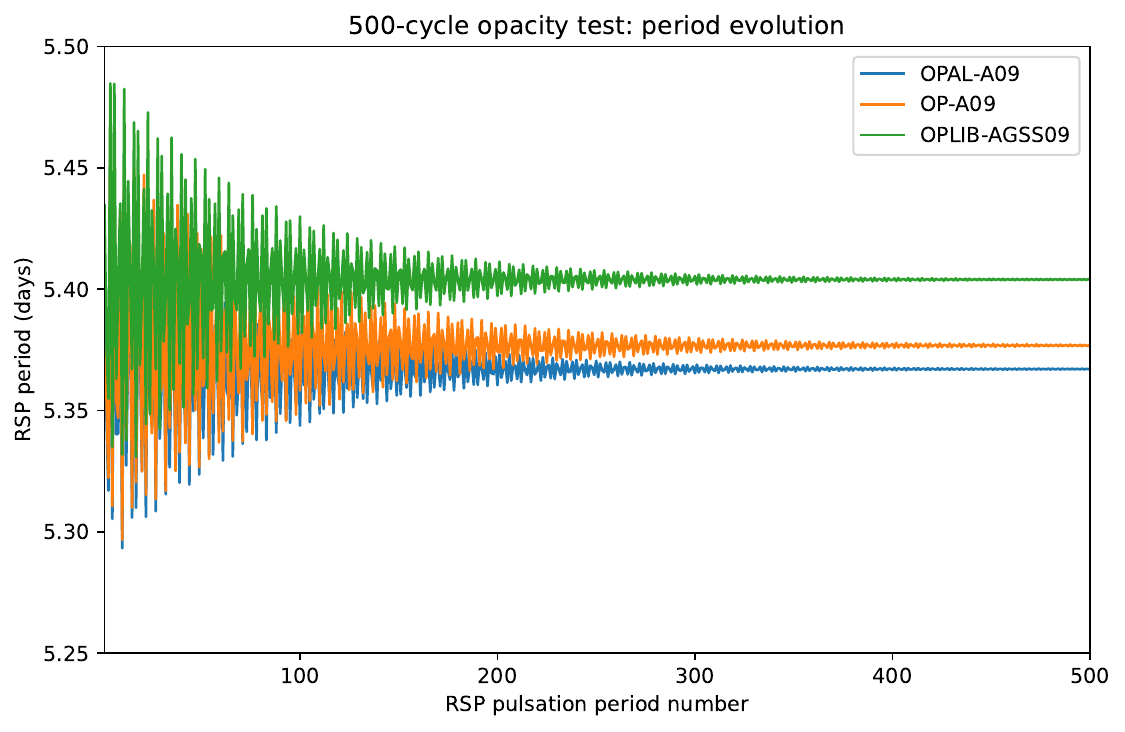}
    \caption{Stitched 1--500-cycle evolution of the MESA-RSP pulsation period for the three native opacity choices, with the initialization record at cycle 0 omitted.  The vertical range is restricted to 5.25--5.50 d to resolve the three nonlinear sequences.  Within this fixed OPAL-calibrated reference configuration, OPAL-A09 is closest to the observed period, while OPLIB-AGSS09 produces a systematically longer period.}
    \label{fig:period_evolution}
\end{figure}

Figure~\ref{fig:period_evolution} shows the stitched 1--500-cycle evolution of the RSP period for the three native opacity cases.  The \texttt{rsp\_num\_periods}=0 history records are omitted from all evolution figures because they belong to model initialization rather than to a completed nonlinear cycle.  In the raw histories those period-zero records contain values near 3.8 d and \GREKM{}=0; the run logs show that nonlinear period measurement begins only after the first radius maximum, with first measured periods of 5.38895, 5.40198, and 5.43460 d for OPAL-A09, OP-A09, and OPLIB-AGSS09, respectively.  Within the fixed OPAL-calibrated reference configuration, OPAL-A09 remains closest to the observed period of \dcep.  At 500 cycles, the OPAL-A09 model gives
\begin{equation}
    P_{\rm RSP}=5.366986\,{\rm d},
\end{equation}
which differs from the adopted observed period by
\begin{equation}
    P_{\rm RSP}-P_{\rm obs}=0.000455\,{\rm d}\simeq 39\,{\rm s}.
\end{equation}
By comparison, OP-A09 gives \(P_{\rm RSP}=5.377089\,{\rm d}\), about 15.2 min longer than the observed period, while OPLIB-AGSS09 gives \(P_{\rm RSP}=5.403926\,{\rm d}\), about 53.8 min longer than the observed period.

This result shows that the opacity choice changes the final nonlinear pulsation period even though the stellar parameters are fixed.  The difference is small in fractional terms but significant for a model that is calibrated against a precisely known Cepheid period.

\subsection{Amplitude-growth diagnostics}

\begin{figure}[t]
    \centering
    \includegraphics[width=0.86\linewidth]{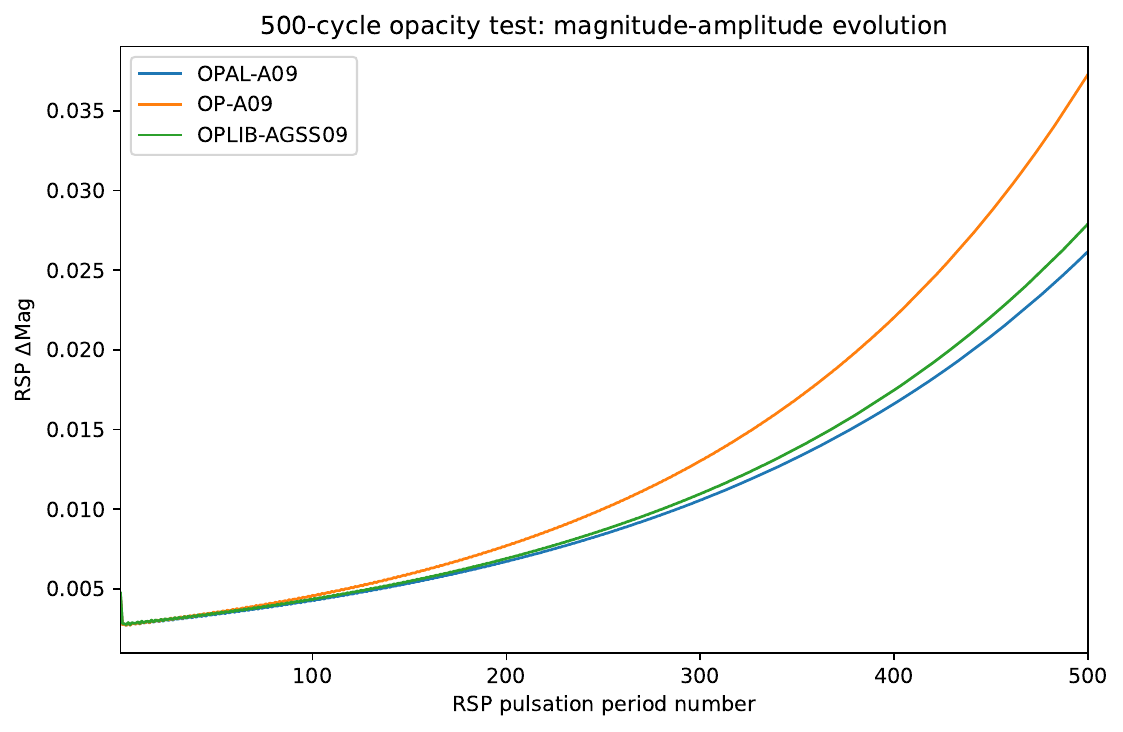}
    \caption{Stitched 1--500-cycle evolution of the MESA-RSP magnitude-amplitude diagnostic \DMag{}, with the initialization record omitted.  OP-A09 produces the largest amplitude-growth response.  The plotted quantity is an internal RSP diagnostic and is not a calibrated Johnson-\(V\) amplitude.}
    \label{fig:deltamag_evolution}
\end{figure}

\begin{figure}[t]
    \centering
    \includegraphics[width=0.86\linewidth]{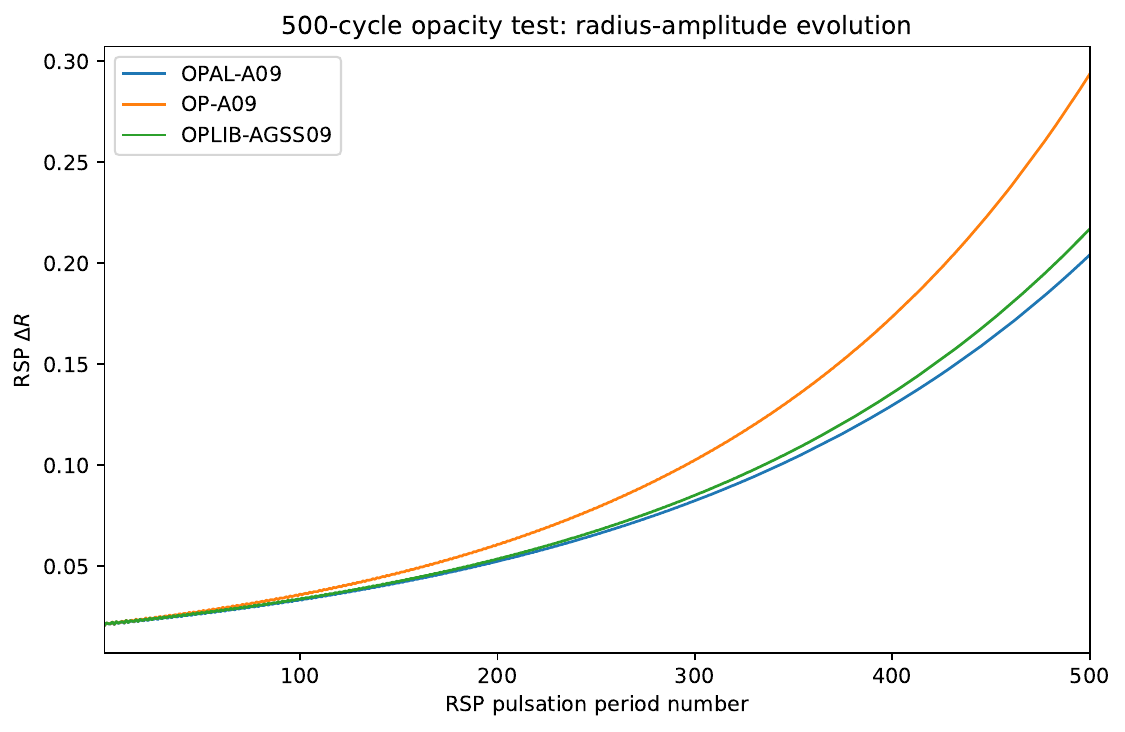}
    \caption{Stitched 1--500-cycle evolution of the MESA-RSP radius-amplitude diagnostic \DR{}, with the initialization record omitted.  OP-A09 gives the largest radius-amplitude growth by 500 cycles.}
    \label{fig:deltaR_evolution}
\end{figure}

Figures~\ref{fig:deltamag_evolution} and \ref{fig:deltaR_evolution} show the evolution of \DMag{} and \DR{}.  At early times the amplitude diagnostics are similar, but by 300--500 cycles the OP-A09 case grows more strongly than the OPAL-A09 and OPLIB-AGSS09 cases.

The final 500-cycle values are summarized in Table~\ref{tab:rsp_500_summary}.  OP-A09 gives \(\Delta{\rm Mag}=0.037307\) and \(\Delta R=0.293677\).  Relative to OPAL-A09, these correspond to increases of 42.5\% and 43.9\%, respectively.  OPLIB-AGSS09 gives smaller increases of 6.6\% and 6.3\% relative to OPAL-A09.

As an additional amplitude constraint, the final complete pulsation cycle gives photospheric radial-velocity proxies \DVphot{} of 1.76, 2.53, and 1.86 \({\rm km\,s^{-1}}\) for OPAL-A09, OP-A09, and OPLIB-AGSS09, respectively.  These values are obtained by finite-differencing \texttt{photosphere\_r}(t) in the MESA history output.  They preserve the same qualitative amplitude ordering as \DMag{} and \DR{}.  They are much smaller than the independent observed spectroscopic peak-to-peak amplitude \(38.9\pm4.9\,{\rm km\,s^{-1}}\) reported by \citet{dasilva2022}; however, the model derivative should be regarded only as a photospheric velocity proxy because no projection factor or spectral-line formation calculation is applied.

\subsection{\texorpdfstring{The \texttt{rsp\_GREKM} nonlinear growth-rate diagnostic}{The rsp-GREKM nonlinear growth-rate diagnostic}}

\begin{figure}[t]
    \centering
    \includegraphics[width=0.86\linewidth]{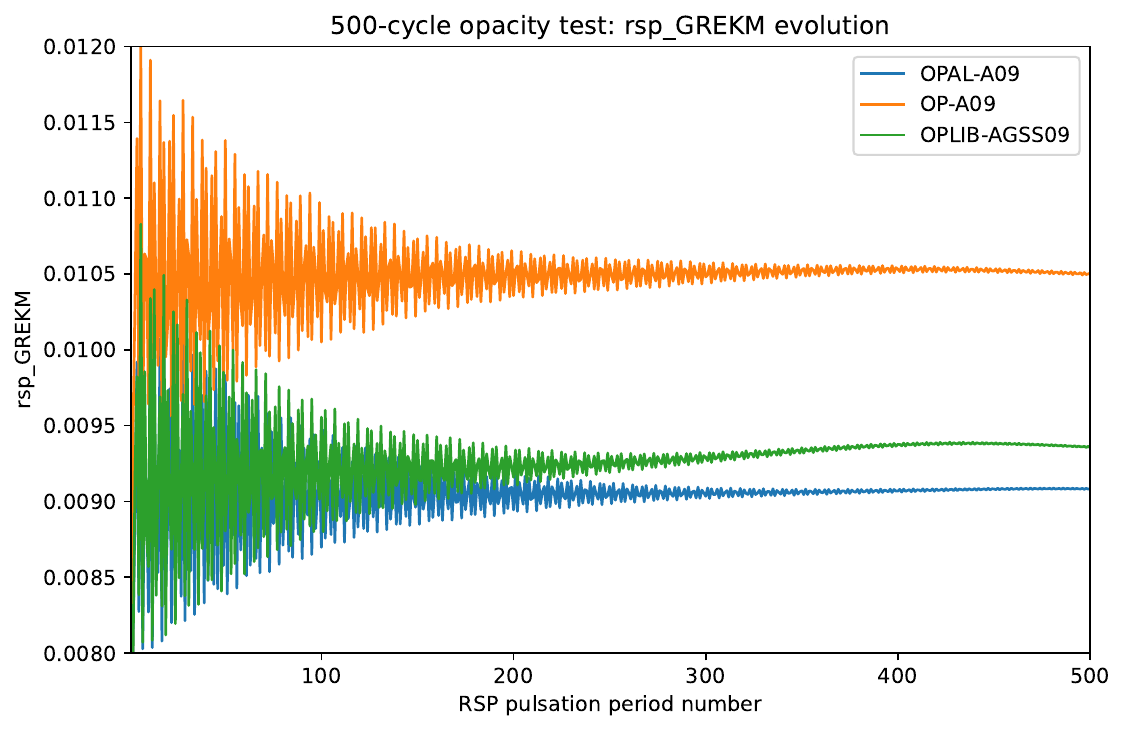}
    \caption{Stitched 1--500-cycle evolution of the internal MESA-RSP \GREKM{} nonlinear growth-rate diagnostic.  The period-zero initialization value \GREKM{}=0 is omitted, and the vertical range is restricted to 0.008--0.012 to resolve the three sequences.  \GREKM{} is used only as a secondary relative diagnostic.  OP-A09 has the largest final \GREKM{} value, consistent with its larger amplitude-growth diagnostics.}
    \label{fig:grekm_evolution}
\end{figure}

Figure~\ref{fig:grekm_evolution} shows the MESA history-column diagnostic \GREKM{}.  The MESA r24.08.1 default history-column list defines this quantity as the fractional growth of kinetic energy per pulsation period, or nonlinear growth rate \citep{mesa_history_columns_2024}.  It is not treated as an observable and is not used as the primary result of the paper.  Its value here is that it gives an independent internal indication of relative pulsation growth among the opacity cases.  The final ordering agrees with the amplitude diagnostics: OP-A09 gives the largest \GREKM{}, OPLIB-AGSS09 is intermediate, and OPAL-A09 is lowest.  This consistency supports the interpretation that the stronger \DMag{} and \DR{} response in OP-A09 is connected to stronger model pulsation growth rather than to an isolated output-column fluctuation.

\subsection{Final diagnostic tables}

Tables~\ref{tab:rsp_500_summary} and \ref{tab:rsp_500_relative} separate the absolute 500-cycle diagnostics from their differences relative to OPAL-A09.  The period shifts are sub-percent (0.188\% for OP and 0.688\% for OPLIB), whereas the OP-A09 amplitude response is much larger: \DMag{} and \DR{} increase by 42.5\% and 43.9\%, and \GREKM{} increases by 15.6\%.  The added photospheric-velocity proxy follows the same amplitude ordering.  These tables therefore quantify both the modest period sensitivity and the stronger opacity dependence of the nonlinear amplitude-growth diagnostics.

\begin{table}[t]
\centering
\caption{Final 500-cycle MESA-RSP opacity-comparison diagnostics.  \GREKM{} is included as a secondary internal growth-rate diagnostic for relative comparison among opacity cases.  The photospheric-velocity proxy is the peak-to-peak finite-difference derivative of \texttt{photosphere\_r} over the final complete cycle.}
\label{tab:rsp_500_summary}
\begin{adjustbox}{max width=\textwidth}
\begin{tabular}{lrrrrrrrr}
\toprule
Opacity case & \(P_{\rm RSP}\) (d) & \(\Delta{\rm Mag}\) & \(\Delta R\) & \GREKM{} & \(\Delta v_{\rm phot,proxy}\) & \(\log T_{\rm eff}\) & \(\log L\) & \(\log R\) \\
 & & & & & (km s\(^{-1}\)) & & & \\
\midrule
OPAL-A09 & 5.366986 & 0.026181 & 0.204153 & 0.009083 & 1.756 & 3.781616 & 3.371846 & 1.648574 \\
OP-A09 & 5.377089 & 0.037307 & 0.293677 & 0.010504 & 2.527 & 3.781311 & 3.371467 & 1.649027 \\
OPLIB-AGSS09 & 5.403926 & 0.027913 & 0.216953 & 0.009356 & 1.858 & 3.781531 & 3.371689 & 1.648681 \\
\bottomrule
\end{tabular}
\end{adjustbox}
\end{table}

\begin{table}[t]
\centering
\caption{Relative 500-cycle differences with respect to OPAL-A09.  The \GREKM{} column gives the relative change in the internal MESA-RSP growth-rate diagnostic.}
\label{tab:rsp_500_relative}
\begin{adjustbox}{max width=\textwidth}
\begin{tabular}{lrrrr}
\toprule
Opacity case & \(\Delta P/P_{\rm OPAL}\) (\%) & \(\Delta(\Delta{\rm Mag})\) (\%) & \(\Delta(\Delta R)\) (\%) & \(\Delta\GREKM{}\) (\%) \\
\midrule
OPAL-A09 & 0.000 & 0.000 & 0.000 & 0.000 \\
OP-A09 & 0.188 & 42.495 & 43.851 & 15.649 \\
OPLIB-AGSS09 & 0.688 & 6.616 & 6.270 & 3.010 \\
\bottomrule
\end{tabular}
\end{adjustbox}
\end{table}

\subsection{Run-length behavior}

Table~\ref{tab:runlength_summary} summarizes the final values at 20, 100, 300, and 500 cycles.  The early 20-cycle values are best interpreted as verification outputs, not as mature nonlinear amplitudes.  The longer integrations show that the opacity-dependent ordering becomes more distinct with increasing cycle number.  
Within the fixed OPAL-calibrated reference configuration, OPAL-A09 converges toward the closest period match, OP-A09 develops the strongest amplitude-growth diagnostics, and OPLIB-AGSS09 remains the longest-period case.

\begin{table}[t]
\centering
\caption{Run-length summary for the native-opacity comparison.  The 20-cycle rows verify successful case-specific opacity-table selection and RSP initialization; no opacity table is switched during a run.  The 500-cycle rows provide the main production comparison.}
\label{tab:runlength_summary}
\begin{adjustbox}{max width=\textwidth}
\begin{tabular}{rlrrrr}
\toprule
Cycles & Opacity case & \(P_{\rm RSP}\) (d) & \(\Delta{\rm Mag}\) & \(\Delta R\) & \GREKM{} \\
\midrule
20 & OPAL-A09 & 5.325074 & 0.003001 & 0.023259 & 0.008869 \\
20 & OP-A09 & 5.324854 & 0.003034 & 0.023737 & 0.010163 \\
20 & OPLIB-AGSS09 & 5.400352 & 0.003000 & 0.023095 & 0.009732 \\
100 & OPAL-A09 & 5.343734 & 0.004303 & 0.033586 & 0.008698 \\
100 & OP-A09 & 5.358233 & 0.004595 & 0.036262 & 0.010051 \\
100 & OPLIB-AGSS09 & 5.429967 & 0.004318 & 0.033578 & 0.009556 \\
300 & OPAL-A09 & 5.365350 & 0.010567 & 0.082463 & 0.009034 \\
300 & OP-A09 & 5.377282 & 0.013026 & 0.102607 & 0.010480 \\
300 & OPLIB-AGSS09 & 5.405368 & 0.010959 & 0.085124 & 0.009291 \\
500 & OPAL-A09 & 5.366986 & 0.026181 & 0.204153 & 0.009083 \\
500 & OP-A09 & 5.377089 & 0.037307 & 0.293677 & 0.010504 \\
500 & OPLIB-AGSS09 & 5.403926 & 0.027913 & 0.216953 & 0.009356 \\
\bottomrule
\end{tabular}
\end{adjustbox}
\end{table}

\section{Discussion}
\label{sec:discussion}

\subsection{A period--amplitude tradeoff among opacity cases}

The controlled comparison shows a useful distinction between period matching and amplitude-growth behavior.  Within the OPAL-calibrated fixed reference model, OPAL-A09 gives the closest period agreement with \dcep{}, but it does not give the largest amplitude-growth diagnostics.  OP-A09 gives the strongest amplitude-growth response, but its final period is farther from the observed value.  OPLIB-AGSS09 shifts the period most strongly while producing only modest amplitude-growth increases relative to OPAL-A09.

This means that opacity choice does not move every diagnostic in the same favorable direction.  In this fixed model, the opacity table that gives the closest period is not the one that gives the largest amplitude growth.  This is important because a Cepheid model calibrated only to the period could still carry a different opacity-dependent amplitude response.  Conversely, a change that increases the amplitude-growth diagnostics may degrade the period match unless the stellar and RSP parameters are retuned.  Because the reference parameters themselves were obtained in an OPAL-based calibration sequence, the present period comparison is not an unbiased competition among independently optimized OPAL, OP, and OPLIB models.  Small changes in \(T_{\rm eff}\), luminosity, mass, or convection parameters could plausibly bring an OP- or OPLIB-based model into comparably close period agreement.  We deliberately do not perform that retuning here because it would change more than the opacity table and therefore broaden the scope beyond the controlled sensitivity experiment.

\subsection{Implications for the amplitude discrepancy}

The observed Johnson-\(V\) amplitude of \dcep{} is approximately \(0.8390\,{\rm mag}\), while the largest internal RSP magnitude-amplitude diagnostic in this experiment is \(\Delta{\rm Mag}=0.037307\) for OP-A09.  Because \DMag{} is not a calibrated Johnson-\(V\) amplitude, this is not a direct one-to-one comparison.  Nevertheless, the difference in scale shows that native opacity-table selection alone is not a complete solution to the model amplitude problem in this fixed setup.  
The same conclusion is independently supported by the final-cycle photospheric-velocity proxies, which span only 1.76--2.53 \({\rm km\,s^{-1}}\) peak to peak, compared with the observed spectroscopic value \(38.9\pm4.9\,{\rm km\,s^{-1}}\) from \citet{dasilva2022}.  The model velocity is not a direct synthetic spectroscopic observable, but its small scale provides a bandpass-independent indication of the remaining dynamical amplitude mismatch.

The correct interpretation is therefore intermediate.  Opacity matters: it changes period, amplitude-growth diagnostics, and \GREKM{} in a measurable and systematic way.  But opacity choice alone does not remove the need for additional modeling work involving RSP turbulent parameters, stellar position in the instability strip, evolutionary state, and synthetic photometry.  The present result is best viewed as a controlled baseline for those broader studies.

\subsection{Paths toward an improved observed-amplitude match}
\label{sec:future_amplitude}

The small photospheric-velocity proxies and the internal magnitude-amplitude diagnostics indicate that none of the present models reproduces the observed dynamical amplitude. There are several physically motivated ways to investigate this mismatch, but they are hypotheses for new calculations, not results of the present opacity comparison. First, the RSP turbulent-viscosity coefficient \texttt{RSP\_alfam} controls eddy-viscous dissipation: testing values below the fixed 0.425 reference may increase nonlinear amplitudes, but could also lead to continued growth, irregular behavior, or numerical energy-error warnings. A systematic exploration of the other time-dependent convection controls, including convective and turbulent flux, turbulent pressure, and radiative losses, would therefore be preferable to varying viscosity alone \citep{paxton2019,farag2026cepheid}.

Second, a grid covering \(M\), \(L\), \(T_{\rm eff}\), and an observationally motivated near-solar heavy-element abundance should be examined, rather than assuming that the single OPAL-tuned input set is optimal. Changes in these variables shift the position within the instability strip and the balance of driving and damping; a revised helium/hydrogen mixture may also affect the partial-ionization zones. These changes should be explored separately from, and then jointly with, opacity-table selection. We would first constrain each opacity family using independent period, temperature, radius/luminosity, and composition information, and only then tune the nonlinear convection parameters against the observed pulsation amplitude and light-curve shape.

Finally, an amplitude comparison must use appropriately synthesized Johnson-\(V\) photometry and a physically modeled line-of-sight radial-velocity curve, not equate the internal \DMag{} quantity or a finite-difference photospheric-velocity proxy with an observed bandpass or spectroscopic observable. Candidate models should also be continued long enough to assess genuine limit-cycle saturation, cycle-to-cycle stability, and numerical energy conservation. No such new parameter-grid calculations are claimed here.

\subsection{\texorpdfstring{Role of \texttt{rsp\_GREKM}}{Role of rsp-GREKM}}

\GREKM{} is included because it helps connect the amplitude-growth diagnostics to the internal pulsation-growth behavior.  It is not a primary observable and should not be compared directly with measured properties of \dcep{}.  Its significance in this paper is relative: OP-A09 has the largest \GREKM{} and also the largest \DMag{} and \DR{}.  That agreement supports the interpretation that OP-A09 gives the strongest model pulsation growth among the three opacity cases.

If \GREKM{} had shown a different ordering from \DMag{} and \DR{}, its value would have been mainly diagnostic.  Instead, it reinforces the main amplitude-growth result.  For that reason, it is retained in the tables but described explicitly as a secondary internal MESA-RSP diagnostic defined by the MESA history-column list.

\subsection{Why custom opacity tables are excluded}

The final comparison is restricted to native MESA opacity tables because they provide a reproducible and documented baseline.  Custom-opacity tables, including OPAS-style tests, are scientifically interesting but require a separate validation pipeline.  Such a pipeline must verify file format, abundance mixture, temperature-density coverage, C/O handling, interpolation smoothness, and recovery of a known reference case.  Without these checks, differences between runs could arise from technical table-handling effects rather than physical opacity differences.  Excluding unvalidated custom-opacity runs from the final interpretation makes the present conclusions narrower but more robust.

\subsection{Limitations}

The study is intentionally limited in several ways.  First, it uses a single fixed MESA-RSP model rather than a full grid over mass, luminosity, effective temperature, metallicity, and turbulent-convection parameters.  That fixed model was calibrated in an OPAL-A09 sequence, so the fact that OPAL remains closest to the target period is conditional on the adopted reference parameters and must not be interpreted as a general preference for OPAL.  Second, the 500-cycle integrations are interpreted as a nonlinear opacity-sensitivity comparison, not as proof of full observed-band amplitude saturation.  Third, the internal \DMag{} diagnostic is not a final Johnson-\(V\) light-curve amplitude.  Fourth, the analysis does not include independent nonlinear hydrodynamic pulsation codes.  These limitations do not invalidate the differential opacity result, but they define its scope.

A further limitation is that the inherited \(Z=0.007\) is below a solar-scaled interpretation of the published near-solar spectroscopic metallicity of \dcep{}, and neither \(X\) nor the chosen reference luminosity was determined by a joint observational inference. This does not alter the internal one-variable opacity test, but limits the direct application of its numerical ordering to the real star.  The main value of the present calculation is its controlled design.  By fixing the stellar and RSP parameters and changing only the high-temperature opacity table, the calculation isolates one physical input.  Broader model grids can then determine whether the same opacity ordering persists after retuning the stellar model.

\section{Conclusions}
\label{sec:conclusions}

We have carried out a controlled native-opacity sensitivity experiment for a fixed \dcep{} MESA-RSP model.  The high-temperature opacity source was varied among OPAL-A09, OP-A09, and OPLIB-AGSS09, while the low-temperature opacity prefix, C/O opacity prefix, stellar parameters, and RSP parameters were held fixed.  The C/O Type-2 table is inactive at the adopted hydrogen-rich composition, and each case was constructed directly with its own Type-1 high-temperature opacity table rather than by switching opacity during a run.  The main conclusions are as follows:

\begin{enumerate}
    \item Native MESA high-temperature opacity-table selection produces measurable differences in the nonlinear RSP response of the fixed \dcep{} model.
    \item Within the OPAL-calibrated fixed reference configuration, OPAL-A09 gives the closest 500-cycle period match, \(P_{\rm RSP}=5.366986\,{\rm d}\), about 39 s longer than \(P_{\rm obs}=5.366531\,{\rm d}\).  This does not establish a general preference for OPAL after opacity-specific stellar-parameter retuning.
    \item OP-A09 gives the largest amplitude-growth diagnostics, increasing \DMag{} by 42.5\% and \DR{} by 43.9\% relative to OPAL-A09 at 500 cycles.
    \item OPLIB-AGSS09 gives a systematically longer 500-cycle period, \(P_{\rm RSP}=5.403926\,{\rm d}\), with more modest increases in the amplitude-growth diagnostics.
    \item \GREKM{} follows the same qualitative ordering as the amplitude-growth diagnostics and is retained as a secondary internal MESA-RSP growth diagnostic.
    \item The final-cycle photospheric radial-velocity proxy is also largest for OP-A09 (2.53 \({\rm km\,s^{-1}}\) peak to peak), compared with 1.76 and 1.86 \({\rm km\,s^{-1}}\) for OPAL-A09 and OPLIB-AGSS09.  All remain far below the independent observed spectroscopic amplitude, while recognizing that the model proxy is not a synthetic line-of-sight velocity.
    \item None of the tested native opacity cases resolves the observed optical-amplitude discrepancy by itself.  Opacity-table choice is therefore a measurable contributor to the modeling uncertainty, but not a complete solution.
    \item Custom-opacity and OPAS-style calculations require a separate validation study before being used for physical interpretation.
\end{enumerate}

These results provide a reproducible baseline for future \dcep{} pulsation studies that combine opacity sensitivity with broader RSP parameter calibration, evolutionary-state constraints, and synthetic observed-band light curves.  As described in Section~\ref{sec:future_amplitude}, the next step is an opacity-specific grid in mass, temperature, luminosity, composition, and convection controls, followed by synthetic photometry, projected spectroscopic velocities, and long-run nonlinear convergence tests.

\section*{Data Availability}

The numerical summaries used in the main tables are included in the manuscript.  The MESA inlists, scripts, logs, and figure-generation files can be made available by the corresponding author upon reasonable request and are intended to accompany the dissertation reproducibility archive.

\section*{Acknowledgements}

The authors acknowledge the use of MESA and its Radial Stellar Pulsation module for the calculations described in this work.

\bibliographystyle{plainnat}
\bibliography{delta_cephei_opacity_sensitivity}

\end{document}